\documentclass[11pt,twoside,onecolumn]{article}
\usepackage[]{latexsym,amsmath,amssymb}
\usepackage[]{epsfig}
\newcommand{\be}{\begin{eqnarray}}
\newcommand{\ee}{\end{eqnarray}}

\newcommand{\ve}{\varepsilon}
\newcommand{\lp}{\ell_{\rm p}}
\newcommand{\mpl}{m_{\rm p}}
\newcommand{\gn}{G_{\rm N}}
\newcommand{\rh}{r_{\rm H}}
\newcommand{\Rh}{R_{\rm H}}
\renewcommand{\d}{\mbox{${\rm d}$}}
\newcommand{\ep}{\mathcal{E}_{\rm p}}
\title{\bf Gravitational tests of the Generalized Uncertainty Principle}
\author{Fabio~Scardigli$^{a,b}$\thanks{Corresponding author.
E-mail: fabio@phys.ntu.edu.tw}
$\ $
and
Roberto~Casadio$^c$\thanks{E-mail: casadio@bo.infn.it}
\\
\\
{\em $^a$Dipartimento di Matematica, Politecnico di Milano,}\\
{\em Piazza Leonardo da Vinci 32, 20133 Milano, Italy}\\
{\em $^b$Institute of Physics, Academia Sinica, Taipei 115, Taiwan}
\\
\null
\\
{\em $^c$Dipartimento di Fisica, Universit\`a di
Bologna and I.N.F.N., Sezione di Bologna,}\\
{\em via Irnerio~46, 40126 Bologna, Italy}}
\date{}
\begin{document}
\maketitle
\begin{abstract}
We compute the corrections to the Schwarzschild metric necessary to reproduce the Hawking temperature
derived from a Generalized Uncertainty Principle (GUP), so that the GUP deformation parameter is directly linked to the deformation
of the metric.
Using this modified Schwarzschild metric, we compute corrections to the standard General
Relativistic predictions for the light deflection and perihelion precession, both for planets in the
solar system and for binary pulsars.
This analysis allows us to set bounds for the GUP deformation parameter from well-known astronomical
measurements.
\par
\null
\end{abstract}
\setcounter{equation}{0}
\section{Introduction}
Research on generalizations of the uncertainty principle of quantum mechanics has
nowadays a long history~\cite{GUPearly}.
One of the main lines of investigation focuses on understanding how the Heisenberg
Uncertainty Principle (HUP) should be modified once gravity is taken into account.
Given the pivotal r\^{o}le played by gravitation in these arguments, it is not
surprising that the most relevant modifications to the HUP have been proposed in
string theory, loop quantum gravity, deformed special relativity, and studies of
black hole physics~\cite{VenezGrossMende,MM,kempf,FS,Adler2,SC},
just to mention some of the most notable frameworks.
\par
An interesting novelty, emerged during the last decade or so, is a lively debate
on the \emph{measurable\/} features of various kinds of Generalized Uncertainty
Principles (GUPs).
From more theoretical shores, the discussion has therefore landed on the ground of
experimental predictions about the \emph{size\/} of these modifications, and several
experiments have been proposed to test GUPs in the laboratory.
Among the more elaborated proposals are those, for example, of the groups of
Brukner and Cerdonio~\cite{brukcerd}.
\par
Studies that aim at putting bounds on the deforming parameter of the GUP,
heretofore denoted by $\beta$, date back at least to Brau~\cite{brau},
and can be roughly divided into three different categories
(actually, only two, as we will see).
In the first group one finds papers such as those of
Brau~\cite{brau}, Vagenas~\cite{vagenas}, Nozari~\cite{Nozari},
which use a specific representation of the operators in the deformed
fundamental commutator~\footnote{We shall work with $c=k_B=1$, but explicitly
show the Newton constant $\gn$ and Planck constant $\hbar$.
We also recall that the Planck length is defined as $\lp^2=\gn\,\hbar/c^3$,
the Planck energy as $\ep\,\lp = \hbar\, c /2$, and the Planck mass as
$\mpl=\ep/c^2$, so that $\gn=\lp/2\,\mpl$ and $\hbar=2\,\lp\,\mpl$.}
\be
\left[\hat{X},\hat{P}\right] = i\,\hbar\left(1 + \beta\, \frac{\hat{P}^2}{\mpl^2}\right)
\ ,
\label{[1]}
\ee
in order to compute corrections to quantum mechanical predictions, such as
energy shifts in the spectrum of the hydrogen atom, or to the Lamb shift,
the Landau levels, Scanning Tunneling Microscope, charmonium levels, etc.
The bounds so obtained on $\beta$ are very stringent, but the drawback of
this approach is a potentially strong dependence of the expected shifts on
the specific representation chosen for the variables $X$ and $P$ in the
fundamental commutator~\eqref{[1]}.
\par
In the second group, we can find the works of, e.g., Chang~\cite{LNChang},
Nozari and Pedram~\cite{Nozari2}, where a deformation of classical Newtonian mechanics
is introduced by modifying the standard Poisson brackets in a way that resembles
the quantum commutator,
\be
\left[\hat{x},\hat{p}\right]
=
i\,\hbar\left(1 + \beta_0\, \hat{p}^2\right)
\quad
\Rightarrow
\quad
\{X,P\} = \left(1 + \beta_0\, P^2\right)
\ ,
\ee
where $\beta_0=\beta/\mpl^2$.
In particular, Chang in Ref.~\cite{LNChang} computes the precession of the
perihelion of Mercury directly from this GUP-deformed Newtonian mechanics,
and interprets it as an extra contribution to the well known precession of
$43"/$century due to General Relativity (GR).
He then compares this global result with the observational data,
and the very accurate agreement between the GR prediction and observations
leaves Chang not much room for possible extra contributions to the precession.
In fact, he obtains the tremendously small bound $\beta\lesssim 10^{-66}$.
A problem with this approach is that a GUP-deformed Newtonian mechanics
is simply superposed to the usual GR theory linearly.
One may argue that a modification of GR at order $\beta$ should likewise
be considered, but this is however omitted in Ref.~\cite{LNChang}.
In other words, it is not clear why the two structures, GR and GUP-modified Newtonian
mechanics, should coexist independently, and why the two different precession errors
add into a final single precession angle.
As a matter of fact, in the limit $\beta \to 0$, Ref.~\cite{LNChang} recovers the
Newtonian mechanics, to which GR corrections must be added as an extra
structure.
Clearly, the physical relevance of this approach and the bound that follows
for $\beta$, remain therefore questionable.
\par
Finally, a third group of works on the evaluation of $\beta$ contains, for example, papers
by Ghosh~\cite{ghosh} and Pramanik~\cite{pramanik}.
They use a covariant formalism, first defined in Minkowski space, with the metric
$\eta_{\mu\nu}={\rm diag}(1,-1,-1,-1)$, which can be easily generalized to curved space-times
via the standard procedure $\eta_{\mu\nu} \to g_{\mu\nu}$.
These papers should however be considered as belonging to the second group.
In fact, a closer look reveals that they also start from a deformation of classical Poisson brackets,
although posited in covariant form.
From the deformed covariant Poisson brackets, they obtain interesting consequences,
like a $\beta$-deformed geodesic equation, which leads to a violation of the Equivalence Principle.
They do not deform the field equations or the metric.
In Appendix~\ref{A1}, however, we show that the violation of the Equivalence Principle is
completely due to the deformation of the Poisson brackets, and has nothing to do with the
covariant formalism, or with a deformation of the GR field equations or solutions,
or of the geodesic equation.
Nonetheless, the Ghosh-Pramanik formalism remains covariant when $\beta \to 0$
and reproduces standard GR results in the limit $\beta \to 0$ (this differs, in general, from the
results obtained by papers in the second group).
\par
The novelties of our approach, when compared with the previous ones, are many and
various.
The main point is to start directly from a quantum mechanical effect,
the Hawking evaporation, for which the GUP is necessarily relevant, rather
than postulating specific representations of canonical operators or modifications
of the classical equations of motion.
We connect the deformation of the Schwarzschild metric directly to the
uncertainty relation, without relying on a specific representation of commutators.
We leave the Poisson brackets and classical Newtonian mechanics untouched,
and recover GR, and standard quantum mechanics, in the limit $\beta \to 0$.
In particular, we preserve the Equivalence Principle, and the equation of motion of
a test particle is still given by the standard geodesic equation.
In the present work, this is obtained by deforming a specific solution of the
standard GR field equations, namely the Schwarzschild metric.
In Appendix~\ref{A2}, we display a non-relativistic analog of this procedure.
A further, more profound, step in this direction would be to formulate from scratch
the deformed field equations of GR, not just assume a deformed solution (as we did),
or a deformed kinematics (as Chang, Ghosh, etc.~did).
This task is left for future developments.

%
%
%
\section{Deforming the Schwarzschild metric}
\setcounter{equation}{0}
\label{GUPSchw}
In this section, we start from a known way of deriving the Hawking temperature directly
from the metric of a black hole, and then show how the GUP modifies the Hawking temperature.
These two steps will pave the road to a deformation of the Schwarzschild metric,
constructed so as to reproduce the GUP-modified Hawking temperature.
\subsection{Standard mass-temperature relation}
We consider here a space-time with a metric that locally has the form
\be
\d s^2
=
g_{\mu\nu}dx^\mu dx^\nu
=
F(r)\,\d t^2 -F(r)^{-1}\,\d r^2 -r^2\,\d\Omega^2
\ ,
\label{mo}
\ee
where $\d\Omega^2 = \d\theta^2+\sin^2\theta \,\d\phi^2$.
For the typical cases we shall consider later on, one has
\be
F(r) = 1 -\frac{2\,\gn\,M}{r} + \frac{\gn\,Q^2}{r^2} + \Lambda \,r^2
\ ,
\label{m}
\ee
however, we do not require any specific form for $F(r)$ for the moment.
Note that the time-like coordinate is chosen as $x^0=t$, the parameters $M$ (mass),
$Q$ (electric charge), $\Lambda$ (cosmological constant, up to a factor) are
real and continuous, with $\Lambda<0$ corresponding to a de~Sitter space-time,
and $\Lambda>0$ to an anti~de~Sitter space-time.
The horizons (if any), are located at the positive zeros of the function $F(r)$
(see, for example, Ref.~\cite{43louko}).
\par
We shall loosely follow the derivation in Ref.~\cite{Zee}.
Suppose $r=\rh$ is an horizon, so that $F(\rh)=0$, and consider $r \geq \rh$.
After the Wick rotation $t \to i\,\tau$ the metric reads
\be
\d s^2
=
-\left[F(r)\,\d\tau^2 +F(r)^{-1}\,\d r^2 +r^2\,\d\Omega^2\right]
\ .
\ee
In the region just outside the horizon, $r \gtrsim r_H$, we perform the coordinate
transformation $(\tau, r) \to  (\alpha, R)$ defined by
\be
\left\{ \begin{array}{ll}
R\,\d\alpha = F(r)^{1/2} \,\d\tau
\\
\\
\d R = F(r)^{-1/2} \,\d r
\ ,
\end{array}
\right.
\label{cooT}
\ee
so that the Euclidean metric becomes
\be
ds^2=
-\left[R^2\,\d\alpha^2 +\d R^2 +r^2(R)\,\d\Omega^2\right]
\ .
\label{}
\ee
The first two terms in $\d s^2$ represent the length element squared of flat 2-dimensional
Euclidean space in polar coordinates.
The Euclidean time $\tau$ is therefore proportional to the polar angle $\alpha$.
Now, denote the period of $\tau$ by $\Theta$ (the period of $\alpha$ is of course $2\pi$).
From the second equation we see that $R$ is a function of $r$ only.
Therefore, integrating the first of the Eq.~\eqref{cooT} over a full period, we get
\be
R(r)\int_0^{2\pi} \d\alpha
=
\sqrt{F(r)}\int_0^\Theta \d\tau
\quad
\Longrightarrow
\quad
2\pi\,R(r) = \Theta\,\sqrt{F(r)}
\ .
\label{nearRh}
\ee
We are interested in what happens just outside the horizon,
therefore we can expand $F(r)$ around $\rh$.
Namely, for $(r-\rh)$ small we obtain
\be
\sqrt{F(r)}
=
\left[F(\rh) + F'(\rh)(r-\rh) + \dots\right]^{1/2}
\simeq
\sqrt{F'(\rh)}\,\sqrt{r-\rh}
\ .
\ee
Eq.~\eqref{nearRh} thus becomes
\be
2\pi\,R(r)
\simeq
\Theta\,\sqrt{F'(\rh)}\,\sqrt{r-\rh}
\ .
\label{29}
\ee
The second of the equations~\eqref{cooT} likewise becomes
\be
\d R(r)
\simeq
\frac{\d r}{\sqrt{F'(\rh)}\,\sqrt{r-\rh}}
\ ,
\ee
which yields
\be
R(r)
\simeq
\frac{2}{\sqrt{F'(\rh)}}\,\sqrt{r-\rh}
\ .
\ee
This, together with Eq.~\eqref{29}, implies
\be
\Theta = \frac{4\pi}{F'(\rh)}
\ .
\ee
Now, $\Theta$ is the period of the Euclidean time, which means,
by general principles of QFT, that a quantized scalar field outside
the horizon lives in a heat bath with temperature $T=\hbar\,\Theta^{-1}$.
To conclude, the temperature of the black hole horizon as seen by
a distant observer is in general given by
\be
T = \hbar\,\frac{F'(\rh)}{4\pi}
\ .
\label{Trh}
\ee
In particular, for a Schwarzschild black hole the function $F(r)$
is given by $Q=\Lambda=0$ in Eq.~\eqref{mo} above, the horizon
is at $r_H = 2\,\gn\,M$, and we get
\be
T_{\rm H} = \frac{\hbar}{8\pi\, \gn\,M}
\ ,
\label{Hw0}
\ee
which is the well-known Hawking temperature.
\subsection{GUP modified mass-temperature relation}
The most common form of deformation of the Heisenberg uncertainty relation
(and the form of GUP that we are going to study in this paper) is without doubt
the following
\be
\Delta x\, \Delta p
\geq
\frac{\hbar}{2}\left(1 \ + \ \beta\,\frac{4\,\lp^2}{\hbar^2}\,\Delta p\,^2\right)
=
\frac{\hbar}{2}\left[ 1 + \beta \left(\frac{\Delta p}{\mpl} \right)^2\right]
\ ,
\label{gup}
\ee
which, for mirror-symmetric states (with $\langle \hat{p} \rangle^2 = 0$),
can be equivalently written in terms of commutators as
\be
[\hat{x},\hat{p}]
=
i\hbar \left[1 + \beta \left(\frac{\hat{p}}{\mpl} \right)^2 \right]
\ ,
\ee
since $\Delta x\, \Delta p \geq (1/2) |\langle [\hat{x},\hat{p}] \rangle|$.
The dimensionless parameter $\beta$ is usually assumed to be of order one,
in the most common quantum gravity formulations.
\par
As is well known from the argument of the Heisenberg microscope~\cite{Heisenberg},
the size $\delta x$ of the smallest detail of an object, theoretically detectable
with a beam of photons of energy $E$, is roughly given by
\be
\delta x
\simeq
\frac{\hbar}{2\, E}
\ ,
\label{HS}
\ee
since larger and larger energies are required to explore smaller and smaller details.
From the uncertainty relation~\eqref{gup}, we see that the GUP version of the standard
Heisenberg formula~\eqref{HS} is
\be
\delta x
\simeq
\frac{\hbar}{2\, E}
+ 2\,\beta\,\lp^2\, \frac{E}{\hbar}
\  .
\label{He}
\ee
which relates the (average) wavelength of a photon to its energy $E$.
(The standard dispersion relation $E=p\,c$ is assumed.)
Conversely, with the relation~(\ref{He}) one can compute the energy
$E$ of a photon with a given (average) wavelength $\lambda \simeq \delta x$.
Following the arguments of
Refs.~\cite{FS9506,ACSantiago,CavagliaD,CDM03,Susskind,nouicer,Glimpses},
we can consider an ensemble of unpolarized photons of Hawking radiation
just outside the event horizon of a Schwarzschild black hole.
From a geometrical point of view,
it is easy to see that the position uncertainty of such photons is of
the order of the unmodified Schwarzschild radius
\be
\Rh=2\,\gn\,M
\ .
\label{RH}
\ee
An equivalent argument comes from considering the average wavelength of
the Hawking radiation, which is of the order of the geometrical size
of the hole.
We can estimate the uncertainty in photon position as
\be
\delta x
\simeq
2\,\mu\,\Rh
\ ,
\label{VI.48a}
\ee
where the proportionality constant $\mu$ is of order unity and will be
fixed soon.
According to the equipartition principle, the average energy
$E$ of unpolarized photons of the Hawking radiation is simply
related with their temperature by
\be
E = T
\ .
\label{kt}
\ee
Inserting Eqs.~(\ref{VI.48a}) and (\ref{kt}) into the formula~(\ref{He}),
we have
\be
{4\,\mu\,\gn\, M}
\simeq
\frac{\hbar}{2\,T}
+ 2\,\beta\,\gn\,T
\ .
\label{37}
\ee
\par
In order to fix $\mu$, we consider the semiclassical limit
$\beta \to 0$, and require that formula (\ref{37}) predicts the standard semiclassical
Hawking temperature~\eqref{Hw0}, that is $T(\beta\to 0)=T_{\rm H}$.
This fixes $\mu = \pi$, so that we have
\be
M
=
\frac{\hbar}{8\pi\, \gn\, T}
+\beta\, \frac{T}{2\pi}
\ .
\label{MT}
\ee
This is the mass-temperature relation predicted by the GUP for a Schwarzschild black hole.
Of course this relation can be easily inverted, to get
\be
T
=
\frac{\pi}{\beta}\left(
M-\sqrt{M^2-\frac{\beta}{\pi^2}\,\mpl^2}
\right)
\ ,
\label{TM}
\ee
where we used $\hbar/\gn = 4 \mpl^2$.
Since, however, the term proportional to $\beta$ is small,
especially for solar mass black holes with $M\gg\mpl$,
we can expand in powers of $\beta$, namely
\be
T
=
T_{\rm H} \left(1 + \frac{\beta\,\mpl^2}{4\pi^2\,M^2} + \dots \right)
\ .
\label{Tg}
\ee
To zero order in $\beta$, we recover the usual Hawking formula~(\ref{Hw0}).
Already from here we can extract an interesting estimate of $\beta$, since
the previous expansion is valid only if
\be
\frac{\beta\,\mpl^2}{4\pi^2\, M^2} \ll 1
\ ,
\label{beta}
\ee
which means $\beta \ll 1.3 \cdot 10^{78}$ for a solar mass black hole
with $M\simeq 10^{38}\,\mpl$.
\subsection{GUP modified Schwarzschild metric}
We can legitimately wonder what kind of (deformed) metric would predict a Hawking temperature
like the one inferred from the GUP relation~(\ref{MT}), for a given $\beta$.
Since we are interested only in small corrections to the Hawking formula,
we can consider a deformation of the Schwarzschild metric of the kind
\be
F(r) = 1 - \frac{2 \,\gn\, M}{r} +\ve\, \frac{\gn^{2}\, M^{2}}{r^2}
\ ,
\label{DSCH}
\ee
and we shall look for the lowest order correction in $\ve$.
Note however that, since $\Rh/r\sim10^{-5}$ on the surface of the Sun,
the term proportional to $\ve$ can still be considered small even if $\ve$ is relatively
large.
The horizons are now given by solutions of the equation
\be
r^2 - 2\gn\, M\, r + \ve\, \gn^{2}\, M^2 = 0
\ .
\ee
We choose the root closest to the unmodified Schwarzschild radius~\eqref{RH},
namely
\be
\rh
=
\Rh\,\frac{1 + \sqrt{1 - \ve}}{2}
\  ,
\ee
which is valid for $\ve \leq 1$ (and possibly negative).
Then
\be
F'(r)
=
\frac{2\,\gn\, M}{r^2} -2 \,\ve\, \frac{\gn^2\,M^2}{r^3}
\ee
and
\be
F'(\rh)
=
\frac{2}{\gn\, M}\, \frac{\sqrt{1-\ve}}{\left( \ 1 + \sqrt{1-\ve} \ \right)^2}
\simeq
\frac{1}{\Rh} \left(1-\frac{\ve^2}{16} + \dots\right)
\ ,
\ee
where the last expansion is valid for $|\ve | \ll 1$.
Hence, the temperature predicted by this deformed Schwarzschild metric is
\be
T(\ve)
=
\hbar\,
\frac{F'(\rh)}{4\pi}
=
\frac{\hbar}{2\pi\, \gn\, M}\, \frac{\sqrt{1-\ve}}{\left( \ 1 + \sqrt{1-\ve} \ \right)^2}
\ ,
\label{Ts}
\ee
which must coincides with the temperature $T(\beta)$ predicted by
Eq.~(\ref{TM}), for any given $\beta$.
That is, $T(\ve)$ should solve Eq.~(\ref{MT}),
\be
M = \frac{\hbar}{8\pi\, \gn\, T(\ve)} +\frac{\beta}{2\pi}\,T(\ve)
\ ,
\ee
which yields a relation between $\beta$ and $\ve$,
\be
\beta(\ve) = -\pi^2 \,\frac{\gn\, M^2}{\hbar}\, \frac{\ve^2}{1-\ve}
\ .
\label{be}
\ee
For $|\ve| \ll 1$, to the lowest order in $\ve$, we thus get
\be
\beta = - \pi^2\,\ve^2\,\frac{M^2}{4\,\mpl^2}
\ ,
\label{be1}
\ee
where we recall that both $\beta$ and $\ve$ are dimensionless.
It is now of great interest to note that Eq.~(\ref{be}) forces us to admit that
$\beta < 0$, since $\ve \leq 1$.
Although quite unexpected, this is a suggestion of fundamental importance.
It seems that a metric is able to reproduce the GUP-deformed Hawking
temperature \emph{only if\/} the deforming parameter $\beta$ is \emph{negative\/}.
We already encountered a situation like this when we studied the uncertainty
relation formulated on a crystal lattice~\cite{JKS}.
This could be a further hint that the physical space-time has actually
a lattice or granular structure at the level of the Planck scale.
\section{Light deflection by deformed Schwarzschild metric}
\setcounter{equation}{0}
\label{GUPlight}
\begin{figure}[t]
\centering
\epsfxsize=16cm
\epsfbox{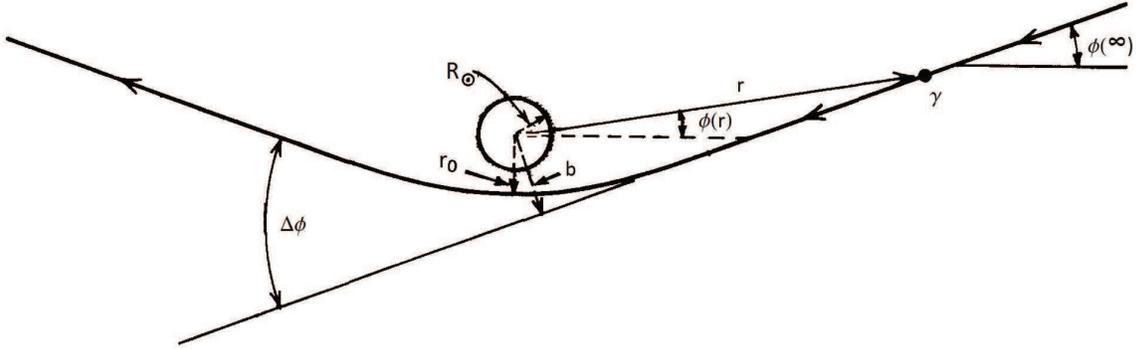}
\caption{Deflection of light by the Sun (refer to the text).
\label{fig1}}
\end{figure}
Having established a connection between the GUP parameter $\beta$ and
the deformation $\ve$ of the Schwarzschild metric, we are now in a position
to compute the physical (possible observable) consequences of such a
deformed metric.
To begin with, we examine the unbound orbits around a massive body,
namely the light deflection by the Sun.
Our treatment roughly follows that of Ref.~\cite{Weinberg72}.
\par
With reference to Fig.~\ref{fig1}, we consider a polar coordinate
system centred in the Sun, with $(\phi, r)$ labelling the position of the
incoming photon, $\phi=\phi(r)$ describes the photon orbit,
$R_\odot$ is the radius of the Sun, $r_0$ the minimum distance
between the photon and Sun.
The photon orbit is thus described by
\be
\phi(r) - \phi(\infty)
=
\int_\infty^{r} \frac{1}{r}\left[\left(\frac{r}{r_0}\right)^2 F(r_0) - F(r)\right]^{-1/2}
\d r
\ .
\label{int}
\ee
The global deflection angle of the orbit from a straight line is just twice its change
from $\infty$ to $r_0$, of course minus $\pi$,
\be
\Delta\phi
\equiv
2\left|\phi(r_0) - \phi(\infty)\right| - \pi
\ .
\ee
This integral can be evaluated exactly by using elliptic integrals, which,
however, can only be numerically computed by expanding in a suitable
small parameter.
It is both easier and more useful to expand before integrating.
Care must be taken in choosing the right small parameter, which
will also help in identifying the finite part of the integral.
Here physical intuition comes in help.
In fact, if the central body had a negligible mass, then $F(r) \to 1$
(the Minkowskian limit), and the trajectory of the photon would be a
(almost) straight line.
Departure from a straight line increases as the central mass $M\sim \Rh$
increases, as well as if the minimum distance from the source $r_0$ decreases.
The right parameter by which expanding the integral (\ref{int}) is thus the
ratio $\Rh/r_0$.
\par
Now, the argument of the integral can be written as
\be
\frac{1}{r}\left[\left(\frac{r}{r_0}\right)^2 F(r_0) - F(r)\right]^{-1/2} = \
\frac{1}{r}\left[\left(\frac{r}{r_0}\right)^2  - 1\right]^{-1/2} \times
\left[F(r_0) + \frac{F(r_0) - F(r)}{(r/r_0)^2 - 1}\right]^{-1/2}
\ ,
\ee
and we find, to first order in $\ve$ and to the second order in $\Rh/r_0$,
\be
\left[F(r_0) + \frac{F(r_0) - F(r)}{(r/r_0)^2 - 1}\right]^{-1/2}
&\!\!=\!\!&
\left[1 - \frac{2\,\gn\,M}{r_0} + \ve\,\frac{\gn^{2}\, M^{2}}{r_0^2} -\frac{2\,\gn\,M\,r_0}{r\,(r+r_0)}
+\ve\, \frac{\gn^2\,M^2}{r^2} \right]^{-1/2}
\nonumber
\\
&\!\!\simeq\!\!&
1 + \frac{\gn\,M}{r_0} \ + \ (3-\ve)\frac{\gn^{2} M^{\,2}}{2\, r_0^2} +
\frac{\gn\,M\, r_0}{r\,(r+r_0)}\left(1 + \frac{3\,\gn\,M}{r_0}\right)
\nonumber
\\
&&
-\ve\,\frac{\gn^{2}\,M^{2}}{2\,r^2} + \frac{3\,\gn^2\,M^2\,r_0^2}{2\,r^2\,(r+r_0)^2}
\ ,
\ee
where we employed the expansion $(1+\delta)^{-1/2} \simeq 1-\frac{\delta}{2} \ + \ \frac{3}{8}\,\delta^2$.
The integral~(\ref{int}) now becomes
\be
\phi(r) - \phi(\infty)
\simeq
A + B + C + D
\ ,
\ee
where
\be
A
=
\int_\infty^{r}
\frac{1}{r}\left[\left(\frac{r}{r_0}\right)^2  - 1\right]^{-1/2}
\left[1+ \frac{\gn\,M}{r_0} + (3-\ve) \, \frac{\gn^{2}\, M^{2}}{2\,r_0^2}\right]
\d r
\ ,
\ee
\be
B
=
\int_\infty^{r}
\frac{1}{r}\left[\left(\frac{r}{r_0}\right)^2  - 1\right]^{-1/2} \frac{\gn\,M\,r_0}{r\,(r+r_0)}
\left(1 + \frac{3\,\gn\,M}{r_0}\right) \d r
\ ,
\ee
\be
C
=
-\int_\infty^{r}
\frac{1}{r}\left[\left(\frac{r}{r_0}\right)^2  - 1\right]^{-1/2} \frac{\ve \, \gn^2\,M^2}{2\,r^2}\, \d r
\ee
and
\be
D
=
\int_\infty^{r}
\frac{1}{r}\left[\left(\frac{r}{r_0}\right)^2  - 1\right]^{-1/2} \frac{3\,\gn^{2}\,M^{2}\,r_0^2}{2\,r^2\,(r+r_0)^2}\,
\d r
\ .
\ee
The integrals are all elementary, and we obtain
\be
\phi(r) - \phi(\infty)
&\!\!\simeq\!\!&
-\left[1 +\frac{\gn\,M}{r_0}
+(3-\ve) \, \frac{\gn^{2}\, M^{2}}{2\,r_0^2}\right]
\arcsin\left(\frac{r_0}{r}\right)
\nonumber
\\
&&
+\frac{\gn\,M}{r_0} \left(1 + \frac{3\,\gn\,M}{r_0}\right)
\left[\sqrt{1 - \left(\frac{r_0}{r}\right)^2}
\left( 1 + \frac{r}{r+r_0}\right) + \arcsin\left(\frac{r_0}{r}\right) - 2\right]
\nonumber
\\
&&
-
\frac{\ve \,\gn^2\,M^2}{2\,r_0^2}
\left[\frac{r_0}{2r}\sqrt{1 - \left(\frac{r_0}{r}\right)^2}
- \frac{1}{2}\,\arcsin\left(\frac{r_0}{r} \right)\right]
\nonumber
\\
&&
-
\frac{3 \,\gn^2\, M^2}{2\,r_0^2}
\left[\sqrt{1 - \left(\frac{r_0}{r}\right)^2}
\left(2 - \frac{r_0}{2\,r} - \frac{r^2}{3\,(r+r_0)^2} + \frac{11\, r}{3\,(r+r_0)}\right)
+\frac{7}{2}\,\arcsin\left(\frac{r_0}{r}\right)
-\frac{16}{3}\right]
\nonumber
\ee
and finally
\be
\Delta\phi
\ \simeq \
\frac{2\,\Rh}{r_0}
+
\frac{ \Rh^2}{16\,r_0^2}\left(15\pi - 16 - 3\pi\,\ve\right)
\ ,
\label{deltaphi1}
\ee
which shows that ours is indeed an expansion in $\Rh/r_0$.
Notice also that the term of second order in $\Rh/r_0$ does not vanish when $\ve \to 0$,
since we properly expanded the integrand to this order.
\par
Our result can now be compared with the deflection angle of a light ray
(or a photon) just grazing the Sun surface, which is usually given in the form
\be
\Delta \phi
=
\frac{1}{2} (1+\gamma) \frac{2\,\Rh}{r_0}
\ ,
\label{deltaphi2}
\ee
where $r_0 = R_\odot$, and $\Rh=2\,\gn\,M_\odot$.
The best measurements presently available for the parameter $\gamma$
from the light bending close to the surface of the Sun, are given by the the development
of the very-long-baseline radio interferometry (VLBI, see Ref.~\cite{will}).
A 2004 analysis of almost 2 million VLBI observations of $541$ radio sources,
made by 87 VLBI sites, yielded $\gamma - 1 \simeq (-1.7 \pm 4.5) \cdot 10^{-4}$~\cite{shapiro}.
A 2009 analysis updated to through 2008 data yielded
$\gamma - 1 \simeq (-1.6 \pm 1.5) \cdot 10^{-4}$ \cite{lambert}.
Good measurements were also performed by the optical astrometry satellite
Hipparcos (at the level of 0.1~percent), and significative improvements are
expected from the just launched satellite Gaia (see Ref.~\cite{HipGaia}).
On comparing Eq.~\eqref{deltaphi1} with \eqref{deltaphi2}, we immediately get
\be
\gamma - 1
=
\frac{\gn\,M}{8\,r_0}\left(15\pi-16-3\pi\ve\right)
\ ,
\ee
or
\be
|\gamma - 1|
=
\frac{\gn\,M}{8\,r_0}
\left|15\pi-16-3\pi\,\ve\right|
\lesssim
1.6 \cdot 10^{-4}
\ .
\ee
For $M=M_\odot$ and $r_0=R_\odot$, this means $-60.7 < \ve < 67.3$,
which, together with the mathematical constrain $\ve \leq 1$,
gives a range for the $\ve$ values of about
\be
-65 \lesssim \ve \leq 1
\ .
\ee
Assuming the "worst" situation, namely $\ve \simeq -65$, then Eq.~\eqref{be}
gives the upper bound for the GUP parameter
\be
|\beta|
=
\frac{M^2}{4\,\mpl^2}\,\frac{\pi^2\,\ve^2}{1-\ve}
\lesssim
5.3 \cdot 10^{78}
\ ,
\ee
which is comparable with the limit obtained from general considerations on the validity
of the low $\beta$ expansion~(\ref{beta}).
This not-so-tight bound has to do with the well-known fact that light deflection is not the
most precise test of GR (since light deflection is still a "Newtonian" phenomenon,
see Ref.~\cite{Soldner}).
A much better estimation will be obtained in the next Section, where the perihelion
precession is considered.
\section{Perihelion precession by deformed Schwarzschild metric}
\setcounter{equation}{0}
\label{GUPperi}
\begin{figure}[t]
\centering
\epsfxsize=13cm
\epsfbox{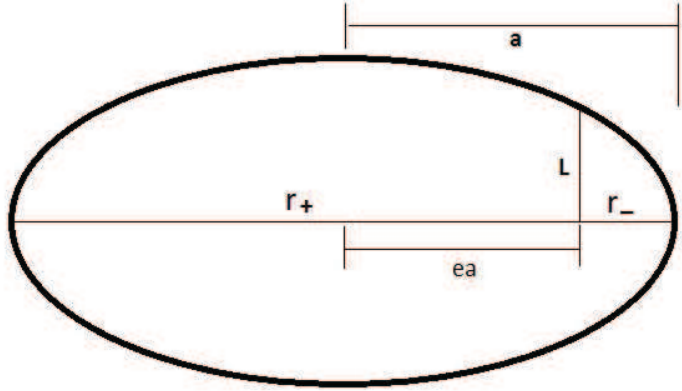}
\caption{Elements of the ellipse used in the text for the calculations of precession of planetary orbits.
\label{fig2}}
\end{figure}
Here, we consider a particle bound in a orbit around a massive body,
typically a planet around the Sun.
Again, we roughly follow the treatment of Ref.~\cite{Weinberg72}.
In Fig.~\ref{fig2} we can see the relevant geometrical parameters for an elliptic orbit
in a polar coordinates system, with the radial coordinate $r$ which at aphelia and
perihelia takes, respectively, the maximum value $r_+$ and minimum value $r_-$;
$e$ is the eccentricity, $a$ the semi-major axis, and $L$ the {\em semilatus rectum\/}.
These geometrical parameters are related by
\be
r_\pm
&\!\!=\!\!&
(1\pm e)\, a
\label{a+-}
\\
L
&\!\!=\!\!&
(1 - e^2)\,a
\label{La}
\\
\frac{2}{L}
&\!\!=\!\!&
\frac{1}{r_+} + \frac{1}{r_-}
\ .
\label{L+-}
\ee
The two relevant constants of motion of the system, $E$ and $J$,
can be interpreted respectively as the energy per unit mass, $(1-E)/2$,
and an angular momentum per unit mass (see Ref.~\cite{Weinberg72}).
The constants $E$ and $J$ can be further expressed as functions of $F(r_-)$
and $F(r_+)$.
The angle swept out by the position vector when it increases from $r_-$ to $r$
is then given by the integral
\be
\phi(r) - \phi(r_-)
=
\int_{r_-}^r
\left[\frac{r_-^2 \left(\frac{1}{F(r)} - \frac{1}{F(r_-)}\right) - r_+^2 \left(\frac{1}{F(r)} - \frac{1}{F(r_+)}\right)}
{r_-^2 r_+^2\left(\frac{1}{F(r_+)} - \frac{1}{F(r_-)}\right)}
- \frac{1}{r^2}\right]^{-1/2}
\frac{\d r}{r^2\,\sqrt{F(r)}}
\ .
\label{traj}
\ee
The total change in $\phi$ at every lap is just twice the change as $r$ increases from
$r_-$ to $r_+$.
This would equal $2\pi$ if the orbit were a closed ellipse, so the total orbital precession
in each revolution is given by
\be
\Delta \phi = 2 \left|\phi(r_+) - \phi(r_-)\right| - 2\pi
\ .
\ee
\par
As before, the exact value of the integral can be expressed via elliptic functions,
but then one has to expand the elliptic functions to obtain useful results.
It is much faster to expand before integrating, where, in analogy with the case of
light deflection, the small parameter is now given by $\Rh/r_-$.
Since $\Rh/r\lesssim\Rh/r_-$ all along the orbit, this also implies that
one can expand in $\Rh/r$.
From the definition of $F(r)$, to second order in $\Rh/r$, we have
\be
\frac{1}{F(r)}
\simeq
1+ \frac{2\,\gn\,M}{r} + (4-\ve)\,\frac{\gn^2\,M^2}{r^2}
\ ,
\ee
so the expression in the square brackets inside the integral in Eq.~\eqref{traj} is actually
a quadratic function of $1/r$.
Moreover, it vanishes for $r=r_\pm$, and we can therefore write
\be
\frac{r_-^2 \left(\frac{1}{F(r)} - \frac{1}{F(r_-)}\right) - r_+^2 \left(\frac{1}{F(r)} - \frac{1}{F(r_+)}\right)}
{r_-^2 r_+^2\left(\frac{1}{F(r_+)} - \frac{1}{F(r_-)}\right)} - \frac{1}{r^2}
=
C \left(\frac{1}{r_-} - \frac{1}{r}\right)\left(\frac{1}{r} - \frac{1}{r_+}\right)
\ ,
\ee
so that the equation~(\ref{traj}) for the trajectory becomes
\be
\phi(r) - \phi(r_-)
=
\frac{1}{\sqrt{C}}
\int_{r_-}^r
\left[\left(\frac{1}{r_-} - \frac{1}{r}\right)\left(\frac{1}{r} - \frac{1}{r_+}\right)  \right]^{-1/2}
\frac{\d r}{r^2\,\sqrt{F(r)}}
\ .
\label{traj2}
\ee
We can find the constant $C$ by considering the limit $r \to \infty$, and hence $F(r)^{-1} \to 1$,
\be
C=
\frac{r_+^2\, F(r_-) \ \left[F(r_+)  -  1\right] \ \ - \ \ r_-^2\, F(r_+) \ \left[F(r_-) - 1\right]}
{r_- r_+ \, [F(r_-) \ - \ F(r_+)]}
\ .
\ee
By inserting the expressions for $F(r_\pm)$ from~(\ref{DSCH}),
and recalling the formula~(\ref{L+-}) for $L$, with a bit of algebra,
we get the exact expression
\be
C
=
\left(1 - \frac{2\,\Rh}{L} + \ve\,\frac{\Rh^2}{L^2} - \ve^2\,\frac{\Rh^3}{8\,a\, L^2}\right)
\left(1 - \ve\, \frac{\Rh}{2\,L}\right)^{-1}
\ee
We can now expand to second order in $\Rh/L$,
\be
C^{-1/2}
\simeq
1 +\left(\frac{4-\ve}{2}\right)\frac{\gn\,M}{L}
+\left(6 -3\ve - \frac{1}{8}\ve^2\right)\frac{\gn^2\,M^2}{L^2}
\ ,
\ee
and, to second order in $\Rh/r$,
\be
[F(r)]^{-1/2}
\simeq
1+\frac{\gn\,M}{r}
+
(3-\ve)\,\frac{\gn^2\,M^2}{2\, r^2}
\ .
\ee
The integral in Eq.~(\ref{traj2}) is largely simplified if we choose a clever change of variable.
Let us consider the closed orbit predicted by Newtonian mechanics.
Taking the pole in the direction of the {\em semilatus rectum\/},
the polar equation of the trajectory reads
\be
\frac{1}{r}
=
\frac{1}{L}-\frac{e}{L}\, \sin\psi
\ ,
\ee
and this will be our change of variable in the integral~(\ref{traj2}),
with
\be
\frac{\d r}{r^2}
=
\frac{e}{L} \cos\psi \, \d\psi
\ .
\ee
On recalling Eq.~\eqref{L+-} and
\be
\frac{e}{L}
=
\frac{1}{2}\left(\frac{1}{r_-} - \frac{1}{r_+}\right)
\ ,
\ee
we have
\be
\frac{1}{r_-}  -  \frac{1}{r}
&\!\!=\!\!&
\frac{e}{L} \ (1+\sin\psi)
\\
\frac{1}{r}  -  \frac{1}{r_+}
&\!\!=\!\!&
\frac{e}{L} \ (1-\sin\psi)
\ ,
\ee
and
\be
\left(\frac{1}{r_-}  -  \frac{1}{r}\right) \left(\frac{1}{r}  -  \frac{1}{r_+}\right)
=
\frac{e^2}{L^2} \cos^2\psi
\ .
\ee
When $r=r_-$, then $\psi=-\pi/2$.
Inserting everything back into Eq.~(\ref{traj2}), to second order in $\Rh/r$, we get
\be
\phi(r) - \phi(r_-)
&\!\!\simeq\!\!&
\left[1+
\left(\frac{4-\ve}{2}\right)\frac{\gn\,M}{L}
+
\left(6 -3\,\ve - \frac{1}{8}\,\ve^2\right)\frac{\gn^2\,M^2}{L^2}
\right]
\nonumber
\\
&&
\times
\int_{-\pi/2}^\psi
\left[ 1 + \frac{\gn\,M}{L}(1-e\,\sin\psi)
+ (3-\ve)\,\frac{\gn^2\,M^2}{2\,L^2}\,(1-e\,\sin\psi)^2
\right]\,
\d\psi
\ .
\ee
At the aphelion $r = r_+$ and $\psi = \pi/2$.
The computation of the integral to the same order in $\Rh/r$ is elementary, and yields
\be
\phi(r_+) - \phi(r_-)
&\!\!\simeq\!\!&
\left[1
+\left(\frac{4-\ve}{2}\right)\frac{\gn\,M}{L}
+\left(6 -3\,\ve - \frac{1}{8}\,\ve^2\right)\frac{\gn^2\,M^2}{L^2}\right]
\nonumber
\\
&&
\times
\left[1+ \frac{\gn\,M}{L}
+(3-\ve)\, \frac{\gn^2\,M^2}{2\,L^2}\left(1+\frac{e^2}{2}\right)
\right]\pi
\nonumber
\\
&\!\!\simeq\!\!&
\pi
\left[1+
\left(\frac{6-\ve}{2}\right)\frac{\gn\,M}{L}
+\frac{\gn^2\,M^2}{L^2}\,N(\ve, e)
\right]
\ ,
\label{expans}
\ee
where
\be
N(\ve,e)
=
\frac{1}{2}\left[19 - 8\,\ve + (3-\ve)\,\frac{e^2}{2} - \frac{\ve^2}{4} \right]
\ .
\ee
Finally the total precession after a single lap is given by
\be
\Delta \phi
\simeq
2\,\pi\left(\frac{6-\ve}{2}\right)\frac{\gn\,M}{L}
+2\,\pi\,\frac{\gn^2\,M^2}{L^2}\,N(\ve, e)
\ .
\ee
In particular, we note that, to first order in $\Rh/L$, we can write
\be
\Delta \phi
\simeq
\frac{6\,\pi\,\gn\,M}{L}\left(1-\frac{\ve}{6}\right)
\ ,
\label{Prec}
\ee
which, of course, reproduces the usual GR prediction in the limit $\ve \to 0$.
This relation should now be compared with known observational data.
\subsection{Solar system data}
The perihelion precession for Mercury is by far the best known and measured
GR precession in the Solar system.
Referring to Ref.~\cite{will} for the latest most accurate and comprehensive data,
we can report the relation
\be
\langle\dot{\omega}\rangle
=
\frac{6\,\pi\, \gn\,M}{L}
\left[\frac{1}{3}(2+2\gamma - \bar{\beta})
+ 3\cdot10^{-4}\,\frac{J_2}{10^{-7}})\right]
\ ,
\ee
where $\langle\dot{\omega}\rangle$ is the measured perihelion shift,
$J_2$ a dimensionless measure of the quadrupole moment of the Sun,
and $\gamma$ and $\bar{\beta}$ are the usual Eddington-Robertson
expansion parameters.
The latest data from helioseismology give $J_2=(2.2\pm0.1) \cdot 10^{-7}$.
The measured perihelion shift of Mercury is known very accurately:
after the perturbing effects of other planets have been accounted for,
the excess shift is known to about $0.1\%$ from radar observations of
Mercury between 1966 and 1990~\cite{Shap2}.
The solar oblateness effect due to the quadrupole moment is then smaller
than the observational error, so it can be neglected.
Substituting standard orbital elements and physical constants for Mercury
and the Sun, we obtain
\be
\langle\dot{\omega}\rangle
=
\left(1 + \frac{2\gamma - \bar{\beta} -1}{3}\right)
\, 42.98"/{\rm century}
\ ,
\ee
where we can place a bound of $|2\,\gamma - \bar{\beta} -1| \lesssim 3 \cdot 10^{-3}$.
\par
Comparing with $\Delta\phi$ from Eq.~(\ref{Prec}), we get
\be
|\ve|
\lesssim
6\cdot 10^{-3}
\ ,
\ee
which, replaced in Eq.~(\ref{be}), yields the lower bound
\be
|\beta|
=
\frac{M^2}{4\,\mpl^2}\,
\frac{\pi^2\,\ve^2}{1-\ve}
\lesssim
3 \cdot 10^{72}
\ .
\label{MS}
\ee
We can also consider the most recent data from the Messenger spacecraft~\cite{laskar},
which orbited Mercury in 2011-2013, and improved very much the knowledge of its orbit.
We can actually push this bound even lower,
to $|2\gamma - \bar{\beta} -1| \lesssim 7.8 \cdot 10^{-5}$,
although the knowledge of $J_2$ would have to improve simultaneously.
If just the error in $|2\gamma - \bar{\beta} -1|$ were taken into account, this would imply
\be
|\ve|
=
2\,\left|2\gamma - \bar{\beta} -1\right|
\lesssim
1.56 \cdot 10^{-4}
\ee
and therefore
\be
|\beta|
\lesssim
2 \cdot 10^{69}
\ .
\ee
But of course this limit should not be considered completely reliable in this contest,
since the less accurate bound on $J_2$ cannot be brutally neglected, at least in principle.
As expected, we gain here at least six orders of magnitude, showing once again that
the perihelion shift is one of the most precise tests of GR, a true GR effect not present
at all in Newtonian gravity (as it is well known).
We can however try to put this limit on a firmer ground  by looking for even larger effects
of this kind.
Where? Of course, in Binary Pulsars!
\subsection{Pulsar PRS B 1913+16 data}
\begin{table}
   \begin{tabular}{ l  l }
     \hline
     Parameter & Value \\ \hline \\
     $e$ (Eccentricity) \dotfill & 0.6171334(5) \\
     $P_b$ (days) (Orbital period) \dotfill & 0.322997448911(4) \\
     $\langle\dot{\omega}\rangle$ (deg/year) (Periastron shift) \dotfill & 4.226598(5) \\
     $\gamma$ (s) (Time dilation-gravitational redshift) \dotfill & $4.2992(8) \times 10^{-3}$ \\
     $\dot{P}_b$ (s/s) (Orbital period decay) \dotfill & $-2.423(1) \times 10^{-12}$ \\ [2mm]
     \hline
   \end{tabular}
   \caption{Orbital parameters of PRS B 1913+16. Figures in parentheses represent estimated uncertainties in the last quoted digit. Data from Ref.\cite{TW}.}
  \label{T1}
\end{table}
Clearly, binary pulsars are very good candidates for measurements of periastron shifts.
Among the known pulsar systems, the best tested pair is the Pulsar PRS B 1913+16.
Discovered in 1974 by Hulse and Taylor, this system has become, after 40 years of
observations, one of the most reliable celestial laboratories for precise GR measurements.
For example, prediction of GR for the period decay rate due to emission of gravitational
waves coincides with the measured value up to an error on the 14th decimal figure.
\par
The state of the art on this system is described in Ref.~\cite{TW}.
In Table~\ref{T1}, we report the orbital parameters of interest for us.
The parameters $e$ and $P_b$ are called Keplerian parameters, since they are well
defined quantities also in the Newtonian theory.
On the contrary, $\langle\dot{\omega}\rangle$, $\gamma$, $\dot{P}_b$ are known as
Damour-Deruelle post-Keplerian parameters~\cite{DD}, quantities typically well defined
in GR only.
In Ref.~\cite{TW89}, Taylor and Weisberg have shown that each GR post-Keplerian
parameter can be expressed in terms of the Keplerian parameters and of the unknown
masses of the pulsar and its companion, $m_1$, $m_2$.
In fact,
\be
\langle\dot{\omega}\rangle
&\!\!=\!\!&
3\, \gn^{2/3}\, c^{-2} (P_b/2\pi)^{-5/3} (1-e^2)^{-1} (m_1 + m_2)^{2/3}
\nonumber
\\
&\!\!=\!\!&
2.113323(2)
\left[\frac{(m_1 + m_2)}{M_\odot}\right]^{2/3} {\rm deg \, \, yr}^{-1}
\label{1}
\\
\gamma
&\!\!=\!\!&
\gn^{2/3}\, c^{-2} \, e \, (P_b/2\pi)^{1/3} \, m_2 \, (m_1 + 2m_2) (m_1 + m_2)^{-4/3}
\nonumber
\\
&\!\!=\!\!&
0.002936679(2)
\left[\frac{m_2 \, (m_1 + 2m_2) (m_1 + m_2)^{-4/3}}{M_\odot^{2/3}} \right] {\rm s}
\label{2}
\\
\dot{P}_b
&\!\!=\!\!&
-\frac{192 \, \pi \, \gn^{5/3}}{5 c^5}\left(\frac{P_b}{2\pi}\right)^{-5/3}
\left(1 + \frac{73}{24}e^2 + \frac{37}{96}e^4\right) (1-e^2)^{-7/2}\, m_1 \, m_2 \, (m_1 + m_2)^{-1/3}
\nonumber
\\
&\!\!=\!\!&
-1.699451(8) \cdot 10^{-12}
\left[\frac{m_1 \, m_2 \, (m_1 + m_2)^{-1/3}}{M_\odot^{5/3}} \right]
\ .
\label{3}
\ee
To compute the numerical coefficients, in the second line of each equation
we have substituted values for $P_b$ and $e$ from Table~\ref{T1},
and used the constants $\gn\,M_\odot/c^3 = 4.925490947 \cdot 10^{-6}\,$s
and 1~Julian~year$\,= 86400 \cdot 365.25\,$s.
The figures in parentheses represent uncertainties in the last quoted digit,
determined by propagating the uncertainties listed in Table~\ref{T1}.
In each case, the errors are dominated by the experimental uncertainty
in orbital eccentricity, $e$.
Furthermore, we note that the analytical expression for the periastron angular
velocity $\langle\dot{\omega}\rangle$ given above by Taylor and Weisberg,
can be obtained from the standard GR prediction for the shift-per-lap $6\pi\,\gn\,M/L$,
simply by dividing this for the period $P_b$, and of course with $M = m_1 + m_2$
(other quantities being expressed as function of $P_b$ and $e$).
\par
To get $\langle\dot{\omega}\rangle^{\rm GR}$, the GR theoretical prediction
of the periastron shift, the strategy is the following:
first, insert the observational values for $\gamma$ and $\dot{P}_b$ from Table~\ref{T1}
into Eqs.~(\ref{2}) and (\ref{3}), and solve for $m_1$, $m_2$.
Then, substitute the values of $m_1$, $m_2$ thus found into Eq.~(\ref{1})
to compute $\langle\dot{\omega}\rangle^{\rm GR}$, and compare this prediction
with the observed value of $\langle\dot{\omega}\rangle^{\rm Obs}$ again
given in Table~\ref{T1}.
The relative error can then be defined as
\be
\tilde{\ve}
=
\frac{\langle\dot{\omega}\rangle^{\rm Obs}-\langle\dot{\omega}\rangle^{\rm GR}}
{\langle\dot{\omega}\rangle^{\rm GR}}
\ ,
\ee
that is $\langle\dot{\omega}\rangle^{\rm GR} (1 + \tilde{\ve}) = \langle\dot{\omega}\rangle^{\rm Obs}$,
and, on comparing with $\Delta\phi$ in Eq.~(\ref{Prec}), we get $|\ve| = 6\,|\tilde{\ve}|$.
\par
There is one further issue we should care of:
measurements are now so precise that the observed value of $\dot{P}_b$ in Table~\ref{T1}
should be corrected for the relative acceleration between the pulsar reference frame and
the solar system centre-of-mass frame (see Ref.~\cite{DT}).
Such relative acceleration is mainly due to the fact that the pulsar and our solar system
are located in different arms of our Galaxy, at different distances from the galactic centre.
The small additional kinematic contribution to the observed $\dot{P}_b$ is
\be
\Delta\dot{P}_{b,gal}
=
\left(-0.027 \pm 0.005\right) \cdot 10^{-12}
\ee
We can solve the system of the two equations (\ref{2}) and (\ref{3}) for the two unknowns
$m_1$ and $m_2$.
The best result is obtained by combining, from Table~\ref{T1}, the lower bound for
$\dot{P}_b = (-2.423-0.001) \cdot 10^{-12}$, and the upper bound of the correction
term $\Delta\dot{P}_{b,gal} = (-0.027 + 0.005) \cdot 10^{-12}$, in order to compute the value
$(\dot{P}_b - \Delta\dot{P}_{b,gal})$ to be inserted in the LHS of Eq.~(\ref{3}).
We then get
\be
\tilde{\ve}
=
8.9 \cdot 10^{-5}
\ ,
\ee
which means $|\ve| \simeq 5.4 \cdot 10^{-4}$.
Recalling now that $M = m_1 + m_2 = 2.828\,\cdot M_{\odot}$, this translates into the bound
\be
|\beta|
\lesssim
2 \cdot 10^{71}
\ ,
\ee
which is tighter than the bound~(\ref{MS}) coming from "standard" Mercury observations,
but weaker than the "Messenger bound" of the previous section.
However, note that we do not have the caveat of the error bounds on $J_2$ here.
\par
Finally, once again, it can be easily checked that the expansion~(\ref{expans})
is fully convergent also when the data of the pulsar PSR~B~1913+16 are inserted,
as well as it is convergent in the solar system field.
\section{Conclusions}
\setcounter{equation}{0}
\label{conc}
We have shown that a suitable deformation of the Schwarzschild metric can reproduce
the Hawking temperature for a black hole, when this is computed from a Generalized
Uncertainty Principle.
We found in this way an analytic relation between the deformation parameter of the metric,
$\ve$, and the usual GUP deformation parameter $\beta$.
In particular, when $\beta \to 0$, we correctly recover GR, and standard quantum mechanics.
Neither the geodesic equation, nor the equivalence principle are violated, for any value of
$\beta$ or $\ve$.
\par
Well-known astronomical measurements, in the Solar system as well as in binary pulsar
systems, allowed us to put constraints on the parameter $\beta$.
This direction seems to point towards promising research:
at present we just deformed the Schwarzschild solution, but a future possibility is to deform
the full field equations of GR, in order to get, among other things, a more stringent bound
on the GUP parameter $\beta$.
We would like to conclude by emphasizing that, although in the existing literature
one can find bounds on $\beta$ much tighter than those obtained in this paper,
they seem to depend, at least partially, either on a specific representation of the deformed
commutator, or on a deformation of Poisson brackets implying a violation of the equivalence
principle.
The line of reasoning followed in the present paper avoids these possible difficulties.
\appendix
\section{Modified Poisson brackets and equivalence principle}
\setcounter{equation}{0}
\label{A1}
We shall show here that a deformation of the classical Newtonian
(i.e.~non covariant) Poisson brackets implies a violation of the
equivalence principle.
In Refs.~\cite{ghosh, pramanik} as well as~\cite{LNChang, Nozari2},
Poisson brackets are deformed in the same fashion as the quantum
commutators derived from a GUP.
\par
Now, considering just a one dimensional system, to keep the calculation
simple, we can write the Poisson brackets for one pair of canonical variables
as
\be
\{q,p\} = 1 + \beta \,p^2 \ ,
\quad \quad \quad
\{q,q\}=\{p,p\}=0
\ .
\label{mP}
\ee
It is then easy to show that for any regular function $H(q,p)$,
the following hold
\be
\{q,H\} = (1 + \beta \,p^2)\frac{\partial H}{\partial p}
\ ,
\quad \quad \quad
\{p,H\} = -(1 + \beta \,p^2)\frac{\partial H}{\partial q}
\ .
\ee
A point-like particle of mass $m$ moving in a Newtonian potential is described by
the Hamiltonian (we assume $M \gg m$)
\be
H = \frac{p^2}{2\,m} - \frac{\gn\,M\,m}{q}
\ ,
\ee
and therefore evolves according to the equations of motion
\be
\dot{q}
&\!\!=\!\!&
\{q,H\}
=(1 + \beta\, p^2)\frac{p}{m}
\nonumber
\\
\label{eqm}
\\
\dot{p}
&\!\!=\!\!&
\{p,H\}
=
-(1 + \beta\, p^2)\frac{\gn\,M\,m}{q}
\ .
\nonumber
\ee
From the first equation, we get
\be
m\,\dot{q}
=
p + \beta\, p^3
\ ,
\label{mdq}
\ee
which implies
\be
m\,\ddot{q}
=
(1 + 3\,\beta\, p^2)\,
\dot{p}
\ ,
\ee
and, using now the second of Eqs.~(\ref{eqm}), we have, to first order in $\beta$,
\be
\ddot{q}
\simeq
- (1 + 4\,\beta\, p^2)\,\frac{\gn\,M}{q^2}
\ .
\ee
Eq.~\eqref{mdq} can be solved for $p$ to the first order in $\beta$, yielding
\be
p
\simeq
m\,\dot{q}
-\frac{\beta(m\,\dot{q})^3}{1 + 3\,\beta\, (m\,\dot{q})^2}
\ .
\ee
Finally, to first order in $\beta$, we have the equation of motion
\be
\ddot{q}
\simeq
- (1 + 4\,\beta\,(m\,\dot{q})^2)\,\frac{\gn\,M}{q^2}
\ .
\ee
Clearly, the trajectory of a test particle of mass $m$ will depend on $m$,
which signals a violation of the Equivalence Principle.
This violation has nothing to do with GR, or the geodesic equation, or the covariant formalism,
but strictly followed from the modified Poisson brackets~\eqref{mP}.
\section{Non-relativistic analogue}
\setcounter{equation}{0}
\label{A2}
We display here a non-relativistic analogue of the deformation procedure followed
in the main text.
For a given metric like the one in Eq.~(\ref{mo}) we know that, in a situation of weak,
stationary field, and for slowly moving particles, we can define the effective (Newtonian)
potential (see, e.g.,~\cite{Weinberg72})
\be
V(r)
=
\frac{1}{2}\,[F(r) - 1]
\ ,
\ee
and the Hamiltonian of a particle of mass $m$, moving in this potential, can be written as
\be
H = \frac{p^2}{2\,m} + m\, V(q)
\ .
\ee
We assume undeformed, standard Poisson brackets
\be
\{q,p\} = 1
\ ,
\quad \quad \quad
\{q,q\} = \{p,p\} = 0
\ ,
\ee
therefore the equations of motion are the usual ones,
\be
\dot{q}
&\!\!=\!\!&
\{q,H\}
= \frac{\partial H}{\partial p} = \frac{p}{m}
\nonumber
\\
\\
\dot{p}
&\!\!=\!\!&
\{p,H\}
=
-\frac{\partial H}{\partial q} = -m\frac{\partial V}{\partial q}
\ ,
\nonumber
\ee
which yield the equation of motion for $q$
\be
\ddot{q}
=
-\frac{\partial V}{\partial q}
\ .
\ee
The above clearly preserves the equivalence principle for any kind of potential
$V(q)$.
In particular, by choosing a deformed metric as in Eq.~(\ref{DSCH}),
we have
\be
\ddot{q}
=
-\frac{\gn\,M}{q^2} + \ve\, \frac{\gn^2\,M^2}{q^3}
\ ,
\ee
which does not depend on $m$ and the equivalence principle is not violated.
\end{document}